\documentclass[conference]{IEEEtran}

\usepackage{cite}
\usepackage{amsmath, amssymb, amsfonts}
\usepackage{graphicx}
\usepackage[table]{xcolor}
\usepackage{url}
\usepackage{booktabs, multirow}
\usepackage{tcolorbox}
\usepackage{float}
\usepackage{balance}

\usepackage{listings}
    \makeatletter
    \def\lst@makecaption{%
      \def\@captype{table}%
      \@makecaption
    }
    \makeatother

\newcommand{\NumberOfManualTestSmells}{eight}
\newcommand{\NumberOfInCompanyTestEngineers}{24}
\newcommand{\AverageApprovalTestEngineers}{80.7\%}
\newcommand{\NumberOfTestSmellOccurrencesExploratoryStudy}{447}
\newcommand{\NumberOfTotalTestSmellOccurrences}{13,169}
\newcommand{\NumberOfTotalTestAllThreeSystems}{2,007}
\newcommand{\SampleSizeExploratoryStudy}{355}
\newcommand{\NLPToolPrecision}{92\%}
\newcommand{\NLPToolRecall}{95\%}
\newcommand{\NLPToolFScore}{93.5\%}

\begin{document}

\IEEEoverridecommandlockouts
\IEEEpubid{\makebox[\columnwidth]{978-1-6654-5223-6/23/\$31.00~\copyright2023~IEEE} \hspace{\columnsep}\makebox[\columnwidth]{ }}

\title{Manual Tests Do Smell! Cataloging and Identifying Natural Language Test Smells}

\author{
    \IEEEauthorblockN{
        Elvys Soares\IEEEauthorrefmark{1}, Manoel Aranda\IEEEauthorrefmark{2}, Naelson Oliveira\IEEEauthorrefmark{2}, Márcio Ribeiro\IEEEauthorrefmark{2}, Rohit Gheyi\IEEEauthorrefmark{3}, Emerson Souza\IEEEauthorrefmark{1},\\Ivan Machado\IEEEauthorrefmark{4}, André Santos\IEEEauthorrefmark{1}, Baldoino Fonseca\IEEEauthorrefmark{2}, and Rodrigo Bonifácio\IEEEauthorrefmark{5}
    }
    \IEEEauthorblockA{
        \IEEEauthorrefmark{1}Universidade Federal de Pernambuco (UFPE), Brazil\\Email: eas5@cin.ufpe.br, epss@cin.ufpe.br, alms@cin.ufpe.br
    }
    \IEEEauthorblockA{
        \IEEEauthorrefmark{2}Federal University of Alagoas (UFAL), Brazil\\Email: mpat@ic.ufal.br, naelson@ic.ufal.br, marcio@ic.ufal.br, baldoino@ic.ufal.br
    }
    \IEEEauthorblockA{
        \IEEEauthorrefmark{3}Federal University of Campina Grande (UFCG), Brazil\\Email: rohit@dsc.ufcg.edu.br
    }
    \IEEEauthorblockA{
        \IEEEauthorrefmark{4}Federal University of Bahia (UFBA), Brazil\\Email: ivan.machado@ufba.br
    }
    \IEEEauthorblockA{
        \IEEEauthorrefmark{5}University of Brasília (UnB), Brazil\\Email: rbonifacio@unb.br
    }
}

\maketitle

\begin{abstract}
Background: Test smells indicate potential problems in the design and implementation of automated software tests that may negatively impact test code maintainability, coverage, and reliability. When poorly described, manual tests written in natural language may suffer from related problems, which enable their analysis from the point of view of test smells. Despite the possible prejudice to manually tested software products, little is known about test smells in manual tests, which results in many open questions regarding their types, frequency, and harm to tests written in natural language. 
Aims: Therefore, this study aims to contribute to a catalog of test smells for manual tests. 
Method: We perform a two-fold empirical strategy. 
First, an exploratory study in manual tests of three systems: the Ubuntu Operational System, the Brazilian Electronic Voting Machine, and the User Interface of a large smartphone manufacturer. We use our findings to propose a catalog of \NumberOfManualTestSmells\ test smells and identification rules based on syntactical and morphological text analysis, validating our catalog with \NumberOfInCompanyTestEngineers\ in-company test engineers. Second, using our proposals, we create a tool based on Natural Language Processing (NLP) to analyze the subject systems’ tests, validating the results.
Results: We observed the occurrence of \NumberOfManualTestSmells\ test smells. A survey of \NumberOfInCompanyTestEngineers\ in-company test professionals showed that \AverageApprovalTestEngineers\ agreed with our catalog definitions and examples. 
Our NLP-based tool achieved a precision of \NLPToolPrecision, recall of \NLPToolRecall, and f-measure of \NLPToolFScore, and its execution evidenced \NumberOfTotalTestSmellOccurrences\ occurrences of our cataloged test smells in the analyzed systems. 
Conclusion: We contribute with a catalog of natural language test smells and novel detection strategies that better explore the capabilities of current NLP mechanisms with promising results and reduced effort to analyze tests written in different idioms.

\end{abstract}


\section{Introduction}
\label{sec:intro}

Test smells are indications of potential problems in the design and implementation of automated software tests~\cite{van2001refactoring}. Like a code smell~\cite{Oliveira2022Lint,medeiros2019investigation}, a test smell does not necessarily mean an already existing problem but an indication of further difficulties such as poor maintainability (\emph{i.e.,} duplication of code~\cite{van2001refactoring}), lack of coverage (\emph{i.e.,} missing or unexecuted verifications~\cite{tahir2016empiricalCat}), or unreliable results (\emph{i.e.,} non-deterministic execution behavior~\cite{palomba2019smell}). 




The necessary investment in configuration can lead a project to opt for manual testing over test automation due to budget limitations~\cite{hauptmann2013hunting, Fernandes2023Put}. In such cases, manual test descriptions are in natural language and \textit{``often of poor quality and written without the best practices of software engineering''}~\cite{hauptmanThesis2016}. Similar to known issues with natural language requirements, documentation of tests in natural language often results in test cases that are incomprehensible, ambiguous, and difficult to maintain due to problems such as translation and spelling errors, different description styles for similar testing procedures, or excessive use of abbreviations~\cite{juhnke2021clustering}.

Despite the format differences, bad choices when implementing automatic tests~\cite{Dalton2020Exception} or describing a manual test using natural language may pose similar threats to the testing activity. For example, Table~\ref{tab:intro:motivating} presents a fragment of a test description from the Ubuntu Operational System (OS) manual tests.\footnote{[Online]. \textit{``testcases\textbackslash{}hardware\textbackslash{}1476\_USB Ports''} test, available: \url{https://git.launchpad.net/ubuntu-manual-tests}} 
In the test, the second action step presents two conditions, \textit{``USB 3.0 storage device''} and \textit{``USB 3.0 port,''} that must be met for the action \textit{``transfer a large
file''} and the corresponding verification step to be performed. The conditional logic phrased in natural language negatively affects test comprehension and correctness. Indeed, as can be seen in Table~\ref{tab:intro:motivating}, a problem in USB 3.0 file transfers may not be identified if the tester does not use compliant equipment and skip step 2. From the point of view of test smells, this is the Conditional Test~\cite{meszaros2007xunit} in natural language~\cite{hauptmann2013hunting}.


\begin{table*}[htbt]
\caption{Steps of An Ubuntu OS test having the Conditional Test smell phrased in natural language}
\label{tab:intro:motivating}
\footnotesize
\centering
\rowcolors{2}{white}{gray!15}
\begin{tabular}{@{}lp{7cm}p{10cm}@{}}
\toprule
\textbf{No} &
  \textbf{Action} &
  \textbf{Verification} \\ \midrule
1 &
  Plug a USB device in and attempt to use it &
  The device is correctly recognized. The software normally used with the device functions normally. The device behaves as expected. The USB device works in every port. You are able to disconnect and re-connect the USB device correctly without errors \\
\textit{2} &
  \textit{If the device is a USB 3.0 storage device and you have a USB 3.0 port, transfer a large file between the two} &
  \textit{The transfer is above USB 2.0 speed} \\
3 &
  Repeat for each USB device you have &
   \\ \bottomrule
\end{tabular}
\end{table*}

Using the rationale presented by the example in Table~\ref{tab:intro:motivating}, Hauptmann \emph{et al.}~\cite{hauptmann2013hunting} coined the term \textit{natural language test smells} to represent possible design problems in manual software testing from the point of view of test smells. A set of seven test smells is presented along with simple detection rules, such as word count or occurrences from keyword lists. After Hauptmann \emph{et al.}~\cite{hauptmann2013hunting} publication, we noticed a research gap of almost ten years concerning natural language test smells, which did not happen in the context of smells in automatic tests~\cite{garousi2018smells, peruma2019distribution, palomba2019smell, panichella2022test}. Such absence motivates a handful of questions concerning the existence of additional natural language test smells, their frequency, the possible problems they indicate, and whether we can benefit from the powerful Natural Language Processing mechanisms available nowadays. These questions motivate our work in this paper, which aims to advance the research on natural language test smells.

We first conduct an exploratory study to analyze a statistically relevant sample of manual test descriptions of three systems from different domains: (i) the Ubuntu Operational System (OS), which is open-source; (ii) the Brazilian Electronic Voting Machine, in an institutional partnership between our institution [name omitted for the blind review process]
and the Superior Electoral Court (TSE); and (iii) a large smartphone manufacturer’s UI --- name omitted due to non-disclosure of proprietary information agreement ---, also in partnership. In this first study, we intend to answer the following research questions:

\begin{itemize}
    \item \textbf{RQ$_{1}$:}\textit{``What already proposed natural language test smells can be observed?''},
    \item \textbf{RQ$_{2}$:}\textit{``What new natural language test smells can be observed?''}, and
    \item \textbf{RQ$_{3}$:}\textit{``How frequent are these test smells?''}
\end{itemize}

Answering these research questions is important to advance the list of test smells applicable to natural language tests. In particular, we identify the occurrence of two already proposed natural language test smells (\emph{i.e.,} \textit{Conditional Test} and \textit{Ambiguous Test}) and contribute to six new smells (\emph{i.e.,} \textit{Unverified Action}, \textit{Misplaced Precondition}, \textit{Misplaced Verification}, \textit{Misplaced Action}, \textit{Eager Action}, and \textit{Tacit Knowledge}), and their frequency in the systems mentioned above. As the final product of this study, we introduce a catalog containing these \NumberOfManualTestSmells\ smells. Our catalog organizes each smell in terms of their name, definition, problem, and identification rules. Instead of using simple detection mechanisms (\emph{e.g.,} searching for a keyword to identify a test smell), our rules are based on powerful natural language processing capabilities like the identification of indefinite determiners, which may indicate non-determinism in the test description.

We conduct an empirical study using an online survey to evaluate our catalog. We recruited \NumberOfInCompanyTestEngineers\ test professionals and presented them with our definitions and examples, asking for their agreement level to our propositions. In this study, we intend to answer the following research question:

\begin{itemize}
    \item \textbf{RQ$_{4}$:}\textit{``How software testing professionals evaluate our proposed smells?''}
\end{itemize}

As a result of our survey, our proposals had an average acceptance of \AverageApprovalTestEngineers\ among the interviewed in-company test engineers, contributing to additional concerns, such as test reproducibility, length, maintainability, and coverage, all originating from doubts raised from poor test writing.

We also contribute to developing an NLP-based tool to identify our catalog's natural language test smells automatically. Our tool implements our defined rules using spaCy,\footnote{[Online]. Available: \url{https://spacy.io/}} a \textit{``free, open-source library for industrial-strength Natural Language Processing (NLP) in Python,''}, and its capabilities concerning syntactic analysis (i.e., elements of the sentences and their properties) like verification verbs and declarative sentences, which are present in multiple languages and whose implementation can be is mostly reused --- as we do to Portuguese, used in the tests of the Brazilian electronic voting machine. To evaluate our tool, we conduct one last empirical study to answer the following research question:

\begin{itemize}
    \item \textbf{RQ$_{5}$:}\textit{``How precise can the automated discovery of natural language test smells be when using NLP?''}
\end{itemize}

The results of this study point to a precision of \NLPToolPrecision, recall of \NLPToolRecall, and f-measure of \NLPToolFScore, indicating a suitable detection level for our proposals. Overall, the tool execution evidenced \NumberOfTotalTestSmellOccurrences\ test smell occurrences in the \NumberOfTotalTestAllThreeSystems\ tests of the analyzed systems, which, by definition, may represent enhancement opportunities to their descriptions.

To sum up, this paper presents the following contributions:

\begin{itemize}
    
    \item We conduct an exploratory study for natural language test smells on systems of different domains: open-source, government, and industry (Section~\ref{sec:exploratoryStudy});
    
    \item We present a catalog of natural language test smells, with six new contributions from our study, along with detection rules that use syntactic and morphological language analysis, representing a novel approach enabled by current NLP technology (Section~\ref{sec:catalog});

    \item We evaluate our catalog with \NumberOfInCompanyTestEngineers\ in-company test engineers (Section~\ref{sec:catalogEvaluation});

    \item We introduce a NLP-based tool to identify the proposed test smells (Section~\ref{sec:tool});

    \item We evaluate our tool by analyzing a sample of its results concerning the before-mentioned systems (Section~\ref{sec:toolEvaluation}).

\end{itemize}

The survey dataset, tool logs, and tool validation records --- for Ubuntu OS tests --- are available online~\cite{Soares2023Replication}. 

\section{Exploratory Study: Towards a Catalog of Natural Language Test Smells}
\label{sec:exploratoryStudy}

This section describes how we analyzed natural language test descriptions to prospect test smells. Also, we give further detail on the selected systems, a sample set of tests, and the distribution (frequency) of our findings. In particular, this exploratory study answers \textbf{RQ$_{1}$}, \textbf{RQ$_{2}$}, and \textbf{RQ$_{3}$}.

\subsection{Planning}
\label{sec:exploratoryStudy:planning}
This exploratory study aims to prospect a set of manual tests from different systems and gather the identified occurrences of test smells. To increase the representativeness of our results, we selected manual tests written in natural language from important systems of three distinct domains: open-source, government, and industry. Considering the limits imposed by the agreements for non-disclosure of confidential information, we detail the obtained tests as follows:

\textbf{\textit{Ubuntu OS:}} As open-source software \cite{ubuntuDownload}, the Ubuntu OS manual tests are available in a public repository.\footnote{[Online]. Available: \url{https://git.launchpad.net/ubuntu-manual-tests}} In the repository, test descriptions are in English and XML format, with standardized tags for test suites, test cases, and action and verification steps. In total, 305 test files containing 973 tests are available.

\textbf{\textit{Brazilian Electronic Voting Machine (BEVM):}} An open-source web-based test management and test execution system manages the manual test descriptions of the BEVM. In the ecosystem, test descriptions are in Portuguese. In total, we had access to 133 tests exported to HTML format.

\textbf{\textit{Large Smartphone Manufacturer (LSM):}} The manual test descriptions of this industry partner are managed by a proprietary issue-tracking product that allows bug tracking and agile project management. Manual test descriptions for this system are in English. In total, 898 test descriptions were made available for our analysis and exported to spreadsheet format.




Three authors manually and independently analyzed a randomly selected subset of test descriptions to perform the exploratory study. Using their know-how on test smells for automatic and manual tests, the authors quoted every questionable description and indicated the possible smell, discussing results in follow-up meetings. It is important to emphasize that access to BEVM and LSM tests was controlled and accessed by cleared authors only. As to the analysis procedure, all authors involved in this activity started with the Ubuntu manual tests to achieve standardization of actions, continuing the analysis in the remaining systems according to their access grants.

Concerning the already proposed smells for tests written in natural language, from the existing list of seven test smells~\cite{hauptmann2013hunting}, five are identified using metrics from an automatic analysis~\cite{sparck1972statistical}: \textit{Badly Structured Test Suite}, \textit{Inconsistent Wording}, \textit{Hard-Coded values}, \textit{Long Test Steps}, and \textit{Test Clones}. As we intended to manually read test descriptions and take notes of the identified problems, using any tool to generate such metrics was out of scope.

Finally, to make our manual analysis effort feasible, we used Cochran's Sample Size Formula~\cite{kotrlik2001organizational} to calculate the sample needed to obtain an 80\% confidence level with a 5\% margin of error for each system individually. Table~\ref{tab:exploring:sampling} presents the analyzed sample test set per system:

\begin{table}[htb]
\caption{Analyzed sample set of tests per system.}
\label{tab:exploring:sampling}
\footnotesize
\centering
\rowcolors{2}{white}{gray!15}
\begin{tabular}{@{}lrr@{}}
\toprule
\textbf{System}                    & \textbf{Manual tests} & \textbf{Sample size} \\ \midrule
Ubuntu OS                          & 973                   & 141                  \\
BEVM                               & 136                   & 75                   \\
LSM                                & 898                   & 139                  \\ \midrule
\multicolumn{1}{r}{\textbf{Total}} & \textbf{\NumberOfTotalTestAllThreeSystems}         & \textbf{\SampleSizeExploratoryStudy}         \\ \bottomrule
\end{tabular}
\end{table}

\subsection{Results}
\label{sec:exploratoryStudy:results}

\begin{figure}[htbp]
  \centering
  \includegraphics[width=.36\linewidth]{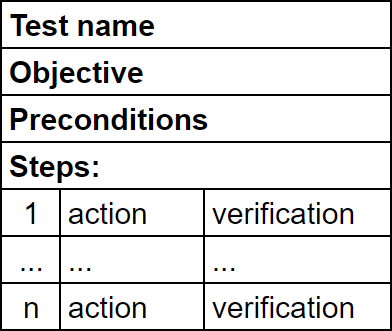}
  \caption{Common test design found in the exploratory study.}
  \label{fig:exploratoryStudy:commonTestDesign}  
\end{figure}

We found similarities in all systems regarding the structure of their manual tests. Although Ubuntu's team does not use a specific test managing tool, they describe their tests as the other two systems, which use open-source and proprietary software for such activities. \figurename~\ref{fig:exploratoryStudy:commonTestDesign} presents a test visualization. Table~\ref{tab:exploratoryStudy:TestStructure} details the test section's writing, regarding the sentence types, with examples from the Ubuntu OS tests.

\begin{table}[htbp]
\caption{Common test structure found in the exploratory study.}
\label{tab:exploratoryStudy:TestStructure}
\footnotesize
\centering
\begin{tabular}{@{}p{1.3cm}p{1.1cm}p{5.55cm}@{}}
\toprule
\textbf{Section}                    & \textbf{Sentence type} & \textbf{Example}                                                        \\ \midrule
\rowcolor{gray!15}
Objective                           & Declarative           & \textit{This test checks that Audio project menu Works}                  \\
\multirow{2}{*}{Preconditions}      & Declarative           & \textit{VMWare Player version $\ge$ 4.0 is required}                     \\
                                    & Imperative            & \textit{Ensure that your system has no Internet access before proceeding} \\
\rowcolor{gray!15}
Action                              & Imperative            & \textit{Click the `Restart now' button}                                         \\
\multirow{2}{*}{Verification} & Declarative & \textit{An `Installation Complete' dialog appears}                                       \\
             & Imperative            & \textit{Verify the system upgraded correctly} \\ \bottomrule
\end{tabular}
\end{table}

\begin{figure}[htbp]
  \centering
  \includegraphics[width=\linewidth]{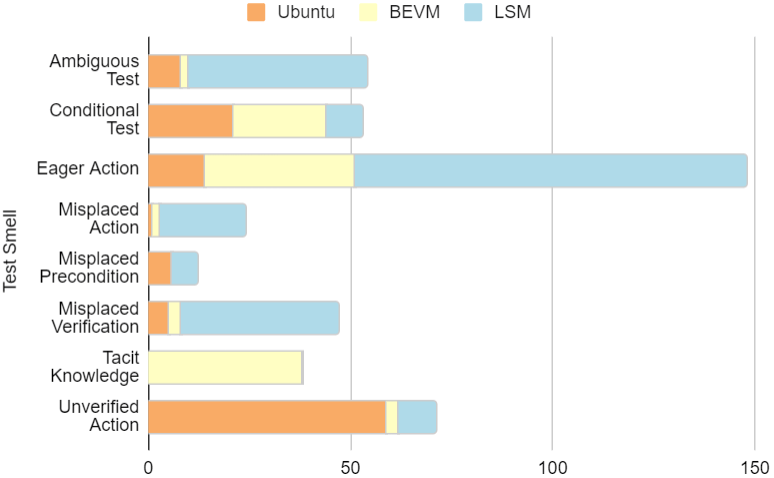}
  \caption{Distribution of identified test smells per system.}
  \label{fig:exploratoryStudy:TestSmellsDistribution}  
\end{figure}

The exploratory study identified \NumberOfManualTestSmells\ test smells, briefly defined in Table~\ref{tab:exploratoryStudy:testSmellDefinitions} and further detailed in Section~\ref{sec:catalog}. From this list, two smells (\emph{i.e.,} \textit{Ambiguous Test} and \textit{Conditional Test}) are proposals from the literature on natural language test smells~\cite{hauptmann2013hunting}, and the remaining ones are contributions from our study. Also, we manually accounted for \NumberOfTestSmellOccurrencesExploratoryStudy\ occurrences of the identified test smells, and \figurename~\ref{fig:exploratoryStudy:TestSmellsDistribution} presents their distribution per system.

\begin{table}[htbp]
\caption{Cataloged test smells}
\label{tab:exploratoryStudy:testSmellDefinitions}
\centering
\footnotesize
\rowcolors{2}{white}{gray!15}
\begin{tabular}{@{}p{2.8cm}p{5.6cm}@{}}
\toprule
\textbf{Test Smell}    & \textbf{Brief definition}                                            \\ \midrule
Ambiguous Test         & Test steps leaving room for interpretation \\
Conditional Test       & Conditional logic phrased in natural language \\
Eager Action           & Single action steps that group multiple actions                \\
Misplaced Action       & Action steps written as verification steps                     \\
Misplaced Precondition & Preconditions as action steps                                  \\
Misplaced Verification & Verification steps written as action steps                     \\
Tacit Knowledge        & Unexplained terms and abbreviations                            \\
Unverified Action      & Action steps without corresponding verifications        \\ \bottomrule
\end{tabular}
\end{table}

\subsection{Discussion}
\label{sec:exploratoryStudy:discussion}

The structural test pattern found in the analyzed systems (Section~\ref{sec:exploratoryStudy:results}) enabled us to propose the test smells presented in Section~\ref{sec:catalog}. Moreover, the distribution of such smells demonstrates the analysis of natural language test descriptions from the point of view of test smells to present promising results. The manual analysis offers some insights whose reality will be precisely shown in Section~\ref{sec:tool}. These insights, for now, indicate that:

\begin{itemize}
    \item Most observed test smells are common to all analyzed systems (\emph{e.g.,} \textit{Eager Action});
    \item There are test smells unique to a single system (\emph{e.g.,} \textit{Tacit Knowledge});
    \item Each system has its own test smell trend (\emph{e.g.,} Ubuntu tests suffer more from \textit{Unverified Action}).
\end{itemize}

\textit{Summary:} Answering \textbf{RQ$_{1}$}, we could observe two already proposed natural language test smells in the analyzed systems. In addition, six new test smells are observed, which answers \textbf{RQ$_{2}$}. Answering \textbf{RQ$_{3}$}, the test smells are frequent throughout the analyzed systems. In particular, Eager Action and Tacit Knowledge tend to be the most and the least frequent ones.

\subsection{Threats to Validity}
\label{sec:exploratoryStudy:threats}

\paragraph{Conclusion} Our identified test smells relate to the common test structure in all three analyzed systems. However, it is important to notice that BEVM and LSM tests are managed by well-adopted software solutions throughout the industry, leading us to understand the found pattern as generally widespread, possibly minimizing this threat.

\paragraph{Internal}
As the accuracy of the exploratory study (\emph{i.e.,} 80\%) is not ideal for generalizations in the analyzed systems (\emph{i.e.,} 95\%~\cite{kotrlik2001organizational}), the distribution of test smells presented in \figurename~\ref{fig:exploratoryStudy:TestSmellsDistribution} may not be precise. We minimized this problem by modeling and validating a Natural Language Processing (NLP)-based tool, further detailed in Section~\ref{sec:tool}, to provide the exact distribution of the presented test smells.

\paragraph{External}
As external threats, analyzing a few software systems may not be enough to identify relevant or well-spread test smells. We minimize this probability by using systems representative of different domains and spoken languages and finding test smells common to such systems.
\section{A Catalog of Natural Language Test Smells}
\label{sec:catalog}

We now present our catalog, the main product of our exploratory study. We show the identified test smells in terms of their names, definition, problem, and identification rules for their detection with examples from the analyzed Ubuntu OS test descriptions.

\subsection{Ambiguous Test}
\label{sec:catalog:AmbiguousTest}

\paragraph{Definition} Originally proposed by Hauptmann \emph{et al.}~\cite{hauptmann2013hunting}, this smell indicates an \textit{``under-specified test that leaves room for interpretation''}. 

\paragraph{Problem} It negatively impacts test comprehension and execution, since the aim needs to be clarified and multiple test executions are not comparable~\cite{hauptmann2013hunting}.

\paragraph{Identification} The original detection rule was the occurrence of any word from a fixed list of ``vague words.''\footnote{similar, better, similarly, worse, having in mind, take into account, take into consideration, clear, easy, strong, good, bad, efficient, useful, significant, adequate, fast, recent, far, close} As Hauptmann \emph{et al.}'s \cite{hauptmann2013hunting} keyword list originated from occurrences in their analyzed test suites and we found a slightly different list in our exploratory study (\emph{e.g.,} some, other, and any), we noticed such keywords to be common in their semantics (syntactic analysis). We propose a more general set of detection rules which consider keyword semantics, and examples, in Table~\ref{tab:catalog:AmbiguousTest}.

\begin{table}[htbp]
\caption{Ambiguous Test Identification}
\label{tab:catalog:AmbiguousTest}
\footnotesize
\centering
\rowcolors{2}{white}{gray!15}
\begin{tabular}{@{}p{3.2cm}p{5.1cm}@{}}
\toprule
\textbf{Rule}  & \textbf{Example} \\ \midrule
Verb + indefinite determiner & \textit{\textbf{Open any} application and suspend machine} \\
Indefinite pronouns      & \textit{At ``Write changes to disks'', verify that \textbf{everything} is right and select YES} \\
Comparative adjectives  & \textit{Is the performance \textbf{similar} or \textbf{better} with no graphical display issues?} \\
Superlative adjectives  & \textit{The root filesystem uses \textbf{most} of the SD card.} \\ 
Adverbs of manner       & \textit{Does fast user switching work \textbf{quickly}?} \\
Comparative adverbs     & \textit{Does everything function \textbf{better} than the stable version?} \\ \bottomrule
\end{tabular}
\end{table}

\subsection{Conditional Test}
\label{sec:catalog:ConditionalTest}

\paragraph{Definition} Tests containing conditional logic phrased in natural language.

\paragraph{Problem} The Conditional Test turns tests very complex and difficult to maintain, negatively impacting test comprehension and correctness since it is hard to understand the intention, and complex tests are more likely to have errors~\cite{hauptmann2013hunting}.

\paragraph{Identification} Originally, Hauptmann \emph{et al.}~\cite{hauptmann2013hunting} proposed a fixed list of words for its detection.\footnote{if, whether, depending, when, in case} As the list is non-exhaustive concerning subordinating conjunctions, we propose any subordinating conjunction, as in Table~\ref{tab:catalog:ConditionalTest}, to identify this smell as a more robust detection rule.

\begin{table}[htbp]
\caption{Conditional Test Identification}
\label{tab:catalog:ConditionalTest}
\footnotesize
\centering
\rowcolors{2}{white}{gray!15}
\begin{tabular}{@{}p{3.2cm}p{5.1cm}@{}}
\toprule
\textbf{Rule}               & \textbf{Example} \\ \midrule
Subordinating conjunctions  & \textit{\textbf{If} you have a USB drive, plug it in.} \\ \bottomrule
\end{tabular}
\end{table}

\subsection{Eager Action}
\label{sec:catalog:EagerAction}

\paragraph{Definition} Single action steps that group multiple actions.

\paragraph{Problem} This test smell may hide implementation problems when any action lacks verification, negatively affecting test effectiveness.

\paragraph{Identification} Imperative verbs represent actions. Example in Table~\ref{tab:catalog:EagerAction}.

\begin{table}[htbp]
\caption{Eager Action Identification}
\label{tab:catalog:EagerAction}
\footnotesize
\centering
\rowcolors{2}{white}{gray!15}
\begin{tabular}{@{}p{2.3cm}p{5.9cm}@{}}
\toprule
\textbf{Rule}             & \textbf{Example} \\ \midrule
Multiple imperative verbs  & \textit{\textbf{Change} some sound settings or other settings (night mode, call history, SMS, etc.) and \textbf{display} them on the phone, \textbf{download} some applications, etc. } \\ \bottomrule
\end{tabular}
\end{table}

\subsection{Misplaced Action}
\label{sec:catalog:MisplacedAction}

\paragraph{Definition} Indicative of a structurally malformed test, the Misplaced Action smell arises when action steps are written as results.

\paragraph{Problem} It negatively impacts test maintainability, since the test structure is not consistent.

\paragraph{Identification} Imperative verbs, excluding verification verbs,\footnote{Verification verbs identified in use: check, verify, observe, recheck} present in verification steps. Example in Table~\ref{tab:catalog:MisplacedAction}.

\begin{table}[htbp]
\caption{Misplaced Action Identification}
\label{tab:catalog:MisplacedAction}
\footnotesize
\centering
\rowcolors{2}{white}{gray!15}
\begin{tabular}{@{}p{3.7cm}p{4.6cm}@{}}
\toprule
\textbf{Rule}      & \textbf{Example} \\ \midrule
Imperatives, excluding verification verbs, as verification steps  & \textit{\textbf{Give} a name to the directory and \textbf{add} files to it as you \textbf{did} in the previous step} \\ \bottomrule
\end{tabular}
\end{table}

\subsection{Misplaced Precondition}
\label{sec:catalog:MispPrecondition}

\paragraph{Definition} Also an indicative of structurally malformed tests, here, preconditions are written as action steps.

\paragraph{Problem} Difficulties in test correctness, since the incorrect placement of preconditions may influence the tester to report test failure should a precondition be unattended.

\paragraph{Identification} When the first action step declares the SUT state. The common format of SUT state is a \textit{noun (subject)} followed by an \textit{auxiliary verb}, followed by a \textit{past participle} verb or adjective in the same sentence (Table~\ref{tab:catalog:MispPrecondition}).

\begin{table}[htbp]
\caption{Misplaced Precondition Identification}
\label{tab:catalog:MispPrecondition}
\footnotesize
\centering
\rowcolors{2}{white}{gray!15}
\begin{tabular}{@{}p{4cm}p{4.2cm}@{}}
\toprule
\textbf{Rule}             & \textbf{Example} \\ \midrule
Subject followed by an auxiliary verb followed by another verb on the past participle & \textit{The \textbf{monitor} \textbf{is} not \textbf{connected}, and the \textbf{PC} \textbf{is} not \textbf{paired}} \\ \bottomrule
\end{tabular}
\end{table}

\subsection{Misplaced Verification}
\label{sec:catalog:MispVerification}

\paragraph{Definition} Another indicative of structurally malformed tests, this smell arises when verification steps written as action steps.

\paragraph{Problem} It negatively impacts test maintainability, since the test structure is not consistent.

\paragraph{Identification} Sentences containing verification verbs written as or along with action steps. Example in Table~\ref{tab:catalog:MispVerification}.

\begin{table}[htbp]
\caption{Misplaced Verification Identification}
\label{tab:catalog:MispVerification}
\footnotesize
\centering
\rowcolors{2}{white}{gray!15}
\begin{tabular}{@{}p{3.9cm}p{4.4cm}@{}}
\toprule
\textbf{Rule}      & \textbf{Example} \\ \midrule
Verification in or as an action step  & \textit{Close flip and \textbf{check} app continuity} \\ \bottomrule
\end{tabular}
\end{table}

\subsection{Tacit Knowledge}
\label{sec:catalog:TacitKnowledge}

\paragraph{Definition} This test smell is related to the use of unexplained terms and abbreviations presuming the tester's familiarity to domain-specific definitions.

\paragraph{Problem} It negatively impacts test comprehension and execution.

\paragraph{Identification} Abbreviations and domain-specific terms not explained in the test description or external reference document (\emph{i.e.,} glossary). A hypothetical example, since we are not authorized to disclose BEVM tests, is in Table~\ref{tab:catalog:TacitKnowledge}.

\begin{table}[htbp]
\caption{Tacit Knowledge Identification}
\label{tab:catalog:TacitKnowledge}
\footnotesize
\centering
\rowcolors{2}{white}{gray!15}
\begin{tabular}{@{}p{4.3cm}p{4cm}@{}}
\toprule
\textbf{Rule}      & \textbf{Example} \\ \midrule
Unexplained terms and abbreviations  & \textit{Check for reported \textbf{residual votes}} \\ \bottomrule
\end{tabular}
\end{table}

\subsection{Unverified Action}
\label{sec:catalog:UnverifiedAction}

\paragraph{Definition} Action steps that miss corresponding verification steps.

\paragraph{Problem} Absent verification steps negatively affect test execution and correctness since there is no instruction on how the system should behave, leaving room for the testers' interpretation.

\paragraph{Identification} Action steps with no corresponding verification steps.
\section{Catalog Evaluation}
\label{sec:catalogEvaluation}

In this section, we present the online survey performed to evaluate our proposals. This activity, in particular, answers \textbf{RQ$_{4}$}.

\subsection{Planning}
\label{sec:catalogEvaluation:planning}

This study planned to assess the opinions of software testing professionals (\emph{e.g.}, engineers, analysts, and managers) about the manual test smells we proposed in Section~\ref{sec:catalog} through an online survey. By stating their agreement with our definitions and examples and commenting on their answers, the software testing professionals would validate whether our proposals represented valid test smells in theory and practice.


We assembled an online survey with questions corresponding to the given definition and example, same as presented in Section~\ref{sec:catalog}. Respondents were presented with the following answering options (unique): \textit{``I strongly agree''}, \textit{``I agree''}, \textit{``Indifferent''}, \textit{``I disagree''}, and \textit{``I strongly disagree''}. Also, every question had an optional comment field.

We recruited participants through individual email invitations, using emails from our industry partner. The invitations were sent to participants of manual test teams, quality assurance professionals, and test managers, none of whom had compensation or obligation to respond to the survey.

\subsection{Results}
\label{sec:catalogEvaluation:results}

\begin{figure*}[htbp]
  \centering
  \includegraphics[width=.95\textwidth]{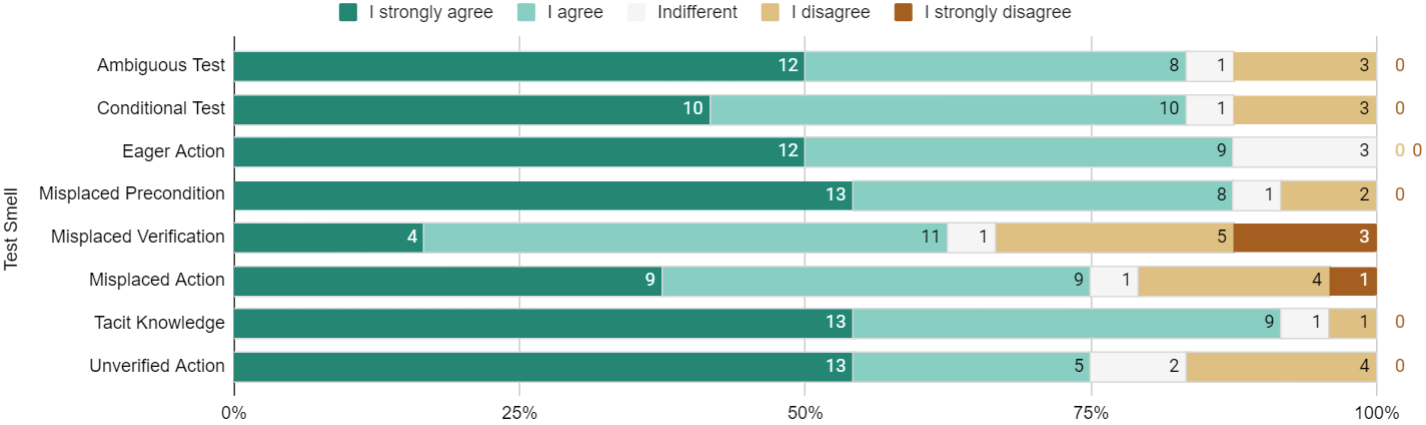}
  \caption{Survey results}
  \label{fig:catalogEvaluation:SurveyResults}
\end{figure*}

We performed the survey in March 2023, achieving \NumberOfInCompanyTestEngineers\ responses for 110 sent emails (21.8\% response rate). Concerning the demographics, we had participants from Brazilian teams, where 83.3\% defined their primary work area as the industry --- over academia --- and their average declared experience with software testing was 4.3 years. \figurename~\ref{fig:catalogEvaluation:SurveyResults} details the results concerning the participants’ opinions on our proposals.

\subsection{Discussion}

Regarding the proposed test smells (Section~\ref{sec:catalog}), the opinion of experienced test professionals (industry partner) served as validation that obtained a high acceptance rate (\figurename~\ref{fig:catalogEvaluation:SurveyResults}). We present the details in the following paragraphs.

Already present in the literature, the \textit{Ambiguous Test} smell (Section~\ref{sec:catalog:AmbiguousTest}) definition was ratified by 83\% of the respondents. Among the agreeing comments, the ambiguity may indeed cause tests to be poorly performed depending on the tester experience, as in \textit{``My experience can improve the test coverage, however for a beginner tester is not be clear the ways to test an interruption, and this can induce he/she to repeat the same procedure/routine or try few different ways to suspend the app.''} Among the testers that disagree with the test smell definition, the variance allowed by non-deterministic terms is beneficial to test different scenarios, as seen in \textit{``I would say that exploratory test cases use a similar approach and it has been working''}.

Known in automatic and manual testing, the \textit{Conditional Test}  (Section~\ref{sec:catalog:ConditionalTest}) smell definition and example had the acceptance of 83\% of the respondents. Concerns about the conditionals being able to improve the test coverage on features not always available arise in both sides, as in the agreeing opinion \textit{``The only part I would not agree is if it is related to a feature that the product may not actually have implemented, for example, NFC.''}, and the disagreeing opinion \textit{``The test writer attempted to cover more possible verifications. If a step or accessory can't be verified all the test is not blocked, and the test becomes applicable to different kinds of product''}.

As the first proposal of our work, the \textit{Eager Action} (Section~\ref{sec:catalog:EagerAction}) test smell definition was ratified by 87.5\% of respondents, with no disagreeing opinion. Among the comments, difficulties in the test execution and concerns about the verifications can be found in \textit{``it seems rather confusing and not pointing to any settings overall, it is covering multiple scenarios''} and in \textit{``There isn't a guarantee that the tester checked all configurations available''}.

Ratified by 75\% of the respondents, the \textit{Misplaced Action} (Section~\ref{sec:catalog:MisplacedAction}) test smell had supporters that manifested concerns about test structure, as in \textit{``Verification steps should be in the end of test cases. Preconditions at the beginning, and actions in the middle.''} Testers that do not agree with the given definition manifest no concern to test structure, since they comprehend the test objective, as in \textit{``If the action keeps the step valid as a single one, it makes sense to be written''}.

The concerns described by the \textit{Misplaced Precondition} (Section~\ref{sec:catalog:MispPrecondition}) test smell definition were accepted by 87.5\% of the respondents. Unfortunately, as that was not a mandatory task in this survey, the respondents provided no comments to this test smell description.

The \textit{Misplaced Verification} (Section~\ref{sec:catalog:MispVerification}) test smell was our least accepted proposal, even though counting with 62.5\% agreeing opinions. Testers claim the test clarity to benefit from the separation into action and verification steps, as in \textit{``I agree because I think that is more clear and organized for the test have it in separated (verification) steps''}. On the opposite hand, testers that did not agree also claimed maintainability benefits of keeping action and verification steps written together, as in \textit{``these actions help to avoid too many steps in a script and reduces the effort in test maintenance''}.

Our most accepted proposal, the \textit{Tacit Knowledge} test smell (Section~\ref{sec:catalog:TacitKnowledge}) definition had the support of 91.6\% respondents. The excessive use of abbreviations and unexplained domain-specific terms is indeed a concern to agreeing respondents, as in \textit{``In my experience, I have faced many new testers and interns having problems knowing abbreviations in test cases''}. Disagreements call attention to test maintainability, as in \textit{``I would say it is a case by case scenario where it could be bad either way. I could have overly long texts due to unnecessary repetition that could be solved by Basic Glossary before the TCs (test cases). Or a inverse scenario where the tester is not provided with edge information to that test.''}

Our last proposal, the \textit{Unverified Action} test smell (Section~\ref{sec:catalog:UnverifiedAction}) had the approval of 75\% respondents. No agreeing respondent gave further details on their answer. Disagreeing respondents manifested concern about the verification steps to every action, as in \textit{``Not every action, in a sequence of actions, generates a relevant result to be verified.''} and \textit{``In some situations the expected result is too obvious and can be dispensed. I believe that this helps to not tire the reader.''}

\textit{Summary:} The online survey shows software testing professionals to mostly agree with our proposals. In addition to positively answering \textbf{RQ$_{4}$}, their comments show additional concerns, such as test reproducibility, length, maintainability, and coverage, all originated from poor test writing.

\subsection{Threats to Validity}

Concerning the internal results, some respondents made the same claim for better organization when a test has action and verification steps written together or separated, for instance, both representing agreeing and disagreeing opinions. However, the wide acceptance of our proposals votes in favor of our interpretation of the possible prejudices, minimizing the threat.

We used responses from software testing professionals who work for our industrial partner, and this bias may influence the generalization of results to other audiences. We minimize this probability through the respondents' experience, of about 4.3 years in average (Section~\ref{sec:catalogEvaluation:results}), and whose answers tend to be similar to experienced professionals who test software in other domains.

Concerning the construct validity, the lack of an attention checking question could bias the results towards the confirmation of our proposals. We minimize this threat by providing examples with attention terms stressed with bold fonts, as in Section~\ref{sec:catalog} examples.
\section{A tool to detect smells in manual tests}
\label{sec:tool}

We present the development of an NLP-based tool, which we call Manual Test Sensei, to detect the natural language test smells we described in Section~\ref{sec:catalog}. This effort shows how implementing our rules for natural language test smells identification is feasible using the current state of the NLP technology.

\begin{figure*}[htbp]
  \centering
  \includegraphics[width=.8\textwidth]{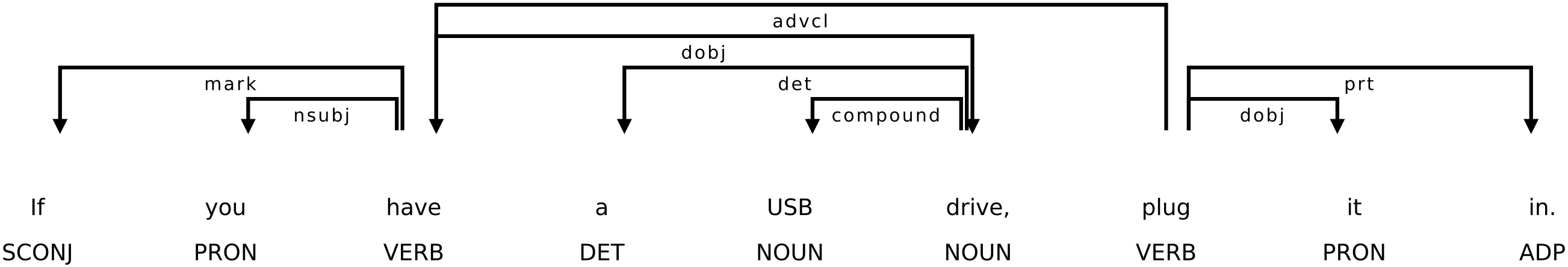}
  \caption{spaCy's visualizer module example}
  \label{fig:tool:DisplacyExample}
\end{figure*}

We use Python and spaCy~\cite{spacy2023}, a commercial open-source software released under the MIT license~\cite{saltzer2020origin}, to implement the NLP tool containing our rules for discovering natural language test smells. SpaCy features convolutional neural network models for part-of-speech (POS) tagging \cite{pennTreebank}, dependency parsing\cite{universalDependencies}, text categorization, and named entity recognition (NER)~\cite{honnibal2017spacy}. \figurename~\ref{fig:tool:DisplacyExample} shows a visualization of the dependency parsing --- arrows above the sentence --- and the POS tagging --- labels beneath each sentence element --- for the Conditional Test example of Section~\ref{sec:catalog:ConditionalTest}. 

The motivation for choosing this combination of programming language and NLP library were (i) using market tools focused on results and performance to analyze industrial-scale software and (ii) the availability of language models beyond English since BEVM tests are in Portuguese.

The chosen strategy enabled us to implement most of our identification proposals. However, identifying  the \textit{Tacit Knowledge} (Section~\ref{sec:catalog:TacitKnowledge}) requires a more comprehensive  solution. To perform it, one would consider (i) external documentation (\emph{e.g.,} glossaries and execution manuals) --- non-existent in Ubuntu and not provided in the BEVM and LSM --- and (ii) a list of standard terms used in manual software testing and considered tacit in every manual testing scenario, where every outsider term would characterize the \textit{Tacit Knowledge} test smell if not clarified. To our knowledge, the proposition of such a list requires a formal study.

\begin{figure}[htbp]
  \centering
  \includegraphics[width=\linewidth]{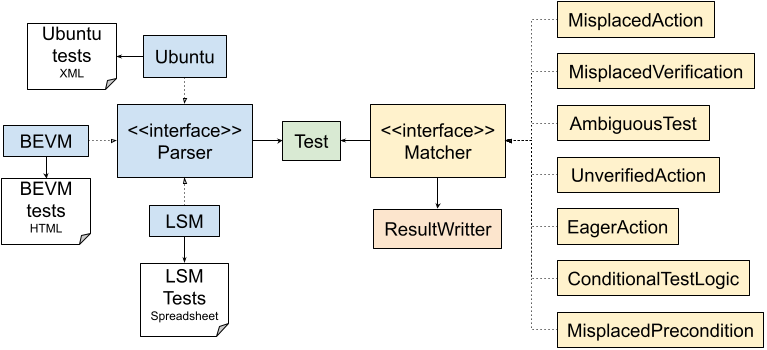}
  \caption{Simplified UML class diagram of the developed NLP tool}
  \label{fig:tool:ClassDiagram}  
\end{figure}

Also, we had to consider the different test file formats according to each analyzed project (Section~\ref{sec:exploratoryStudy:planning}): XML for the Ubuntu OS, HTML for the BEVM tests, and spreadsheet for LSM tests. To that end, specific parsers were created for each system's test file format. \figurename~\ref{fig:tool:ClassDiagram} presents a simplified UML class diagram of the Manual Test Sensei tool, where the parsers --- responsible for transforming a test file into several test objects --- and the test smell matchers are shown. Finally, the tool produces a CSV file as output containing the test file name, the identified test smell, the specific words or sentence span that characterize the test smell, and the analyzed (action or verification) step.

The tool source code is available in an online repository at \url{https://github.com/easy-software-ufal/manual-test-sensei}.
\section{Tool Evaluation}
\label{sec:toolEvaluation}

Once the proposition and development of the Manual Test Sensei tool --- implementing our natural language test smell identification rules --- proved possible using current NLP technology (Section~\ref{sec:tool}), in this last study, we present the tool results and validation, therefore demonstrating how precise is the tool performance. This activity, in particular, answers \textbf{RQ$_{5}$}.

\subsection{Planning}
\label{sec:toolEvaluation:planning}

This study planned to execute the Manual Test Sensei tool against the entire test set of the three analyzed systems and validate the results. Therefore, we could verify whether the distribution found in the exploratory study (Section~\ref{sec:exploratoryStudy}) is maintained in the Manual Test Sensei execution results, as well as the accuracy --- in terms of precision, recall, and f-measure metrics~\cite{van1974foundation, powers2020evaluation} --- of such results.


Although we executed our tool against the entire test set of the three systems, manually validating the tool's output of \NumberOfTotalTestSmellOccurrences\ smells would be infeasible. Therefore, we randomly selected 101 tests distributed in proportion to the number of tests available in every analyzed system. 

For every selected test, an author would first analyze it manually and indicate the found test smells, then verify the tool results for that test, and finally indicate the results that were correct or true positives (TP), incorrect or false positives (FP), and the missed or false negatives (FN) test smells. Table~\ref{tab:toolEvaluation:settings:distributionSampleTests} presents the distribution of the randomly selected tests per system:

\begin{table}[htbt]
\caption{Distribution of selected tests in the validation sample}
\label{tab:toolEvaluation:settings:distributionSampleTests}
\centering
\footnotesize
\rowcolors{2}{white}{gray!15}
\begin{tabular}{@{}lrr@{}}
\toprule
\textbf{System}                    & \textbf{Total tests} & \textbf{Sample size} \\ \midrule
Ubuntu OS                          & 973                  & 49                   \\
BEVM                               & 136                  & 7                    \\
LSM                                & 898                  & 45                   \\ \midrule
\multicolumn{1}{r}{\textbf{Total}} & \textbf{2,007}        & \textbf{101}         \\ \bottomrule
\end{tabular}
\end{table}

\subsection{Results}
\label{sec:toolEvaluation:results}

\begin{figure}[htbp]
  \centering
  \includegraphics[width=\linewidth]{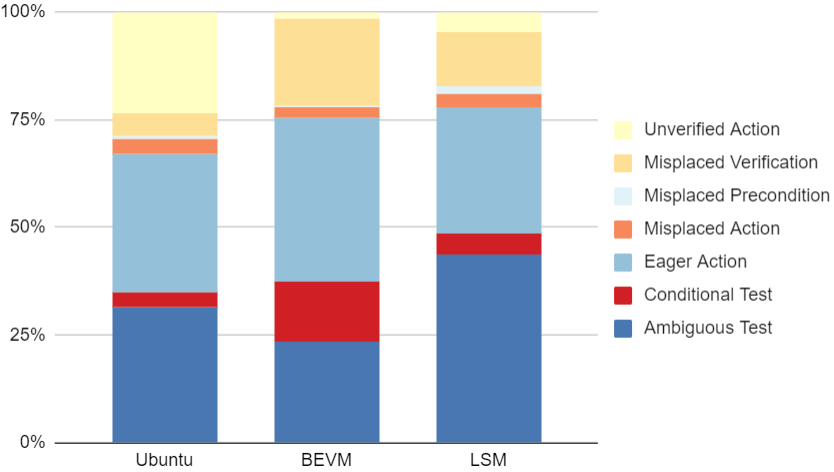}
  \caption{Distribution of test smells per system.}
  \label{fig:toolEvaluation:results:NLPResultsProportional}  
\end{figure}

A total of \NumberOfTotalTestAllThreeSystems\ test descriptions were analyzed by the Manual Test Sensei tool. The tool indicated 13,169 test smells, with an average of 6.5 test smells per analyzed test, noticeably higher than the 1.2 test smells found in the exploratory study (Section~\ref{sec:exploratoryStudy}). Considering the analyzed systems individually, we obtained an average of 8.5 test smells per Ubuntu OS test, 5.8 test smells per BEVM test, and 4.5 test smells per LSM test. Table~\ref{tab:toolEvaluation:results:totalNLPResults} presents the results per test smell and system. Finally, a distribution of the found test smells per analyzed system is presented in \figurename~\ref{fig:toolEvaluation:results:NLPResultsProportional}.

\begin{table}[htbp]
\caption{Total NLP results}
\label{tab:toolEvaluation:results:totalNLPResults}
\centering
\footnotesize
\rowcolors{2}{white}{gray!15}
\begin{tabular}{@{}lrrrr@{}}
\toprule
\textbf{Test Smell}                & \textbf{Ubuntu} & \textbf{BEVM} & \textbf{LSM}   & \textbf{Total}  \\ \midrule
Ambiguous Test                     & 2,627           & 185           & 1,776          & \textbf{4,588}  \\
Conditional Test                   & 277             & 110           & 193            & \textbf{580}    \\
Eager Action                       & 2,664           & 299           & 1,191          & \textbf{4,154}  \\
Misplaced Action                   & 318             & 19            & 124            & \textbf{461}    \\
Misplaced Precondition             & 45              & 3             & 74             & \textbf{122}    \\
Misplaced Verification             & 428             & 161           & 513            & \textbf{1,102}  \\
Unverified Action                  & 1,967           & 11            & 184            & \textbf{2,162}  \\ \midrule
\multicolumn{1}{r}{\textbf{Total}} & \textbf{8,326}  & \textbf{788}  & \textbf{4,055} & \textbf{13,169} \\ \bottomrule
\end{tabular}
\end{table}

Three authors performed the verification as defined in Section~\ref{sec:toolEvaluation:planning}. The selected sample of 101 tests resulted in 708 results for this activity. The analysis of such results by the involved authors resulted in 15 disagreements comprising doubts in syntactical and morphological text analysis, properly clarified in a discussion meeting. Table~\ref{tab:toolEvaluation:results:metrics} presents the detailed validation totals per system and the precision, recall, and f-measure metrics achieved by the tool in this validation activity.

\begin{table}[htbp]
\caption{Detailed NLP tool validation and metrics}
\label{tab:toolEvaluation:results:metrics}
\centering
\footnotesize
\rowcolors{2}{white}{gray!15}
\begin{tabular}{@{}lrrrrrr@{}}
\toprule
\textbf{System}                    & \textbf{TP}  & \textbf{FP} & \textbf{FN} & \textbf{Precision} & \textbf{Recall} & \textbf{F-measure} \\ \midrule
Ubuntu OS                          & 384          & 43          & 18          & 0.9                & 0.96            & 92.64              \\
BEVM                               & 25           & 0           & 0           & 1                  & 1               & 1                  \\
LSM                                & 213          & 13          & 12          & 0.94               & 0.95            & 94.46              \\ \midrule
\multicolumn{1}{r}{\textbf{Total}} & \textbf{622} & \textbf{56} & \textbf{30} & \textbf{0.92}      & \textbf{0.95}   & \textbf{93.53}     \\ \bottomrule
\end{tabular}
\end{table}

\subsection{Discussion}
\label{sec:toolEvaluation:discussion}

The high expressiveness of the adopted technology, either in the identification of dependency relationships (e.g., subject + auxiliary verb + participle verb) or in the identification of the Part of Speech (POS) (e.g., indefinite pronouns), enabled us to implement most of the detection rules as defined in Section~\ref{sec:catalog}. Only one identification rule could not be implemented entirely, which was the Conditional Test, identified through subordinating conjunctions (SCONJ) at the beginning of a dependent clause in a sentence. As spaCy does not natively support splitting sentences into clauses, which varies from language to language, identifying SCONJ in a dependent clause in the middle of a sentence results in many identification problems by the pre-trained models. This problem resulted in 8 false negatives identified in the validation activity, representing approximately 27\% of the test smells not identified by the tool.


We encountered various formatting, spelling, and character encoding conversion issues in the test descriptions. Using numbered and unnumbered lists, parentheses, and the lack of correct punctuation impaired the NLP engine classification in some cases reported as false positives and false negatives. For example, the implemented mechanism was not able to identify a subordinate clause in the sentence \textit{``(If on a `laptop') Is plugged to a power source,''} nor in \textit{``Type in your user name and press Enter (you can accept the default if you wish),''} and could not differentiate the link label in the sentence \textit{``Click the Choose Payment Method link,''} which lacked quotes, and was erroneously classified as multiple actions.

However, even with the implementation challenges and some test malformations mentioned, the result obtained in the metrics of precision, recall, and f-measure for the tool can still be considered expressive. The results remain promising even when using a trained model for a different idiom and executing the same rules --- except for the list of verification verbs used in the \textit{Misplaced Verification} detection, which needed a partner in Portuguese for BEVM tests --- as seen in the metrics presented by Table~\ref{tab:toolEvaluation:results:metrics}. 

According to Table~\ref{tab:toolEvaluation:results:totalNLPResults}, the most frequent test smells detected were the \textit{Ambiguous Test} (\emph{i.e.,} 34.8\%) and \textit{Eager Action} (\emph{i.e.,} 31.5\%). An interesting distribution noticed is that, from the 4,588 occurrences of the \textit{Ambiguous Test}, we accounted for 2,225 (\emph{i.e.,} 48.5\%) occurring in action steps and 2,363 (\emph{i.e.,} 51.5\%) occurrences in verification steps, meaning that ambiguous tests have an almost equal probability of presenting testers with difficulties in \textit{``what to perform in the test''} and \textit{``what to verify as a result.''} However, being less frequent may not mean less harm to the testing activity. It is important to remember that a \textit{Misplaced Precondition} can induce the tester to declare the test failed if the precondition is not met and the test cannot be executed~\cite{soares2020refactoring, soares2022refactoring}.

\begin{figure}[htbp]
  \centering
  \includegraphics[width=\linewidth]{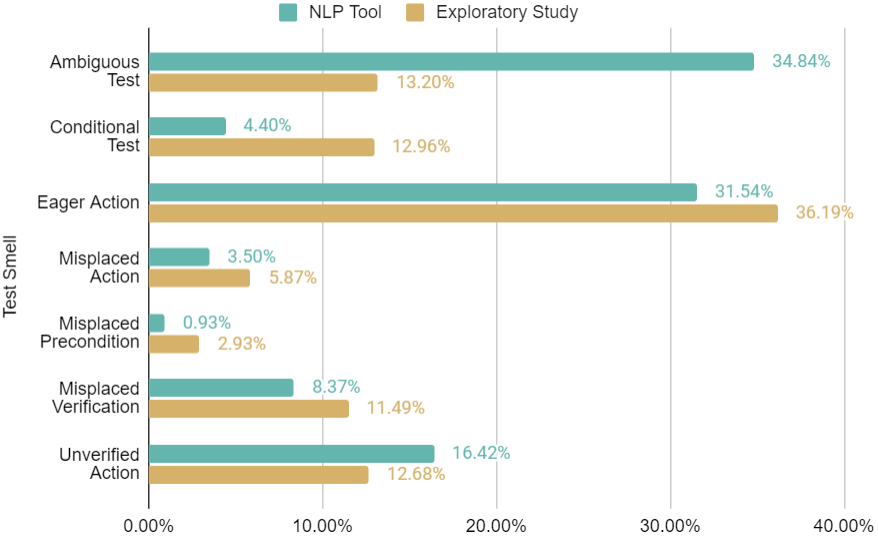}
  \caption{Comparison between the exploratory study and the NLP tool results.}
  \label{fig:toolEvaluation:discussion:comparisonExploratoryNLP}
\end{figure}

Comparing the distribution of test smells found in the exploratory study (Section~\ref{sec:exploratoryStudy}) and the one found by the NLP tool (Section~\ref{sec:toolEvaluation:results}), shown in \figurename~\ref{fig:toolEvaluation:discussion:comparisonExploratoryNLP}, we noticed that some test smells had a different percentage result between the two activities, which was the case of the \textit{Ambiguous Test} and the \textit{Conditional Test}. This expressive difference was due to the more precise identification of the tool in cases of undefined determinants, which may escape the most attentive --- or not sufficiently trained in the exploratory study --- eyes. Still, the precision difference in the exploratory study (Section~\ref{sec:exploratoryStudy}) and the tool validation (Section~\ref{sec:toolEvaluation:results}), necessary for this study to be feasible, influenced the found deviation.

Furthermore, we noticed that test smells not found in the exploratory study for specific systems, such as the \textit{Misplaced Precondition} for the BEVM tests, are now among the results of the NLP tool (Table~\ref{tab:toolEvaluation:results:totalNLPResults}), even with few occurrences (i.e., 3). This result is also expected and included in the exploratory study's 5\% margin of error (Section~\ref{sec:exploratoryStudy:planning}). Finally, the proportional distribution of test smells per system shown in \figurename~\ref{fig:toolEvaluation:results:NLPResultsProportional} shows that although the tests of the analyzed systems suffer from the test smells found, they do so in different proportions.

\textit{Summary:} The results obtained in the tool validation show that our detection rules are effective in identifying the considered test smells. In particular, we achieved a f-measure of 93.53\%, which answers \textbf{RQ$_{5}$}.

\subsection{Threats to Validity}
\label{sec:toolEvaluation:threats}

\textit{Internal:} The tool’s results may contain errors. We manually analyzed 101 tests to minimize this threat, which meant more than 700 results, according to Table~\ref{tab:toolEvaluation:results:metrics}. This amount of results was enough to guarantee statistical validity~\cite{kotrlik2001organizational} for the \NumberOfTotalTestSmellOccurrences\ results generated by the tool.

\textit{External:} The generalization of the obtained results is impossible with the selected sample of three systems. We minimize this threat by choosing highly expressive systems from different domains to analyze. Nevertheless, an exploratory study would confirm whether our results indicate some degree of probability to the analysis of other systems.
\section{Related Work}
\label{sec:related}

Hauptmann \emph{et al.}~\cite{hauptmann2013hunting} presented possible problems in manual test descriptions performed in natural language from the point of view of test smells. Together with coining the term \textit{Natural Language Test Smells}, the authors propose a set of seven smells: \textit{Hard-Coded Values}, \textit{Long Test Steps}, \textit{Conditional Tests}, \textit{Badly Structured Test Suites}, \textit{Test Clones}, \textit{Ambiguous Tests}, and \textit{Inconsistent Wording}. Also, the authors present identification strategies for their proposals that rely on keyword lists and complimentary metrics (\emph{i.e., number of words}) and the frequency of the proposed test smells in nine industrial test suites. In our work, we extend the current catalog by adding six new test smells, their discovery strategies and frequency, and providing updates for the discovery of two of Hauptmann \emph{et al.}'s list, which we base on broader definitions focused on morphological and syntactical language analysis, thus exploring the capabilities of current Natural Language Processing mechanisms.

Rajkovic and Enoiu presented a tool called NALABS to detect bad smells in natural language requirements and test specifications~\cite{NALABS}. Similarly to Hauptmann \emph{et al.}~\cite{hauptmann2013hunting}, the proposed tool uses keyword lists to measure vagueness, referenceability, optionality, subjectivity, and weakness metrics. They also used Automated Readability Index (ARI) to measure readability and the number of words and conjunctions to measure test complexity. Again, our work differentiates from Rajkovic and Enoiu's work because we use current NLP mechanisms to identify words using morphological and syntactical language analysis.

Transferring the concept of code smells to requirements engineering, Femmer \emph{et al.}~\cite{FEMMER2017190} introduced a lightweight static requirements analysis approach that allows for quick checks when requirements are written down in natural language. In another work, Femmer \emph{et al.}~\cite{Femmer2014} derived a set of smells from the natural language criteria of the ISO/IEC/IEEE 29148 standard, showing that lightweight smell analysis can uncover many practically relevant requirements defects. Like our work, they also use tool support to analyze text in natural language descriptions.

Previous works presented test smells in test code. Some of these smells are related to ours, although we focused on natural language test smells.  Meszaros \emph{et al.}~\cite{meszaros2007xunit} and Peruma \emph{et al.}~\cite{peruma2019distribution} studied test smells in test code, such as \emph{Conditional Test} and \emph{Conditional Test Logic}, which are related to Hauptmann \emph{et al.}~\cite{hauptmann2013hunting} natural language test smell. Aljedaani \emph{et al.}~\cite{aljedaaniAssertionless} also listed the \textit{Assertionless Test} smell, defined by the absence of assertions, which is similar to our idea of natural language tests having no verification steps (\textit{Unverified Action}).
\section{Concluding Remarks and Future Work}
\label{sec:conclusion}

In this paper, we extended the current research on Natural Language Test Smells by contributing six new test smell propositions, strategies for their detection, and their frequency in a sample of three representative systems in the government, open-source, and industry domains. Also, we proposed updates for two well-known test smells applicable to natural language test descriptions. Unlike the current research, we proposed a novel detection strategy for natural language test smells that relies on syntactical and morphological text analysis, thus exploring the capabilities of current Natural Language Processing mechanisms.

To conduct this work, we performed two independent and complimentary parts: first, we performed an exploratory study whose results we validated with industry test professionals. We guided the development of an NLP-based tool in the second part. The results and validation metrics showed our strategy to be effective, detecting test smells with over 90\% precision, even in a multiple-idiom context.

In future work, we intend to (i) enable the implementation of the \textit{Tacit Knowledge} test smell by performing a formal study to define common terms in software testing terminology that may be considered tacit in any manual execution of software tests; (ii) execute the tool analysis in other candidate systems whose test management is performed using the same tools as BEVM and LSM tests to verify the generalization of our results; and (iii) aggregate tests --- and test file formats --- from uncovered systems in the results.

Finally, we verified that some cataloged test smells also exist in the case of automatic tests. Moreover, the results of this study show that, like their automatic correlates, natural language test smells are also quite frequent, corroborating the title statement.


\section*{Acknowledgment}
We thank the Brazilian Superior Electoral Court (TSE) and our industrial partner for kindly allowing their tests to be analyzed in our study. This research was partially funded by CNPq grants 312195/2021-4, 421306/2018-1, 310313/2022-8; and FAPEAL grants 60030.0000000462/2020 and 60030.0000000161/2022. Also, this work is partially supported by INES (National Institute of Software Engineering): CNPq grant 465614/2014-0, CAPES grant 88887.136410/2017-00, and FACEPE grants APQ-0399-1.03/17 and PRONEX APQ/0388-1.03/14.

\balance

\bibliographystyle{IEEEtran}
\bibliography{bibliography}

\begin{thebibliography}{10}
\providecommand{\url}[1]{#1}
\csname url@samestyle\endcsname
\providecommand{\newblock}{\relax}
\providecommand{\bibinfo}[2]{#2}
\providecommand{\BIBentrySTDinterwordspacing}{\spaceskip=0pt\relax}
\providecommand{\BIBentryALTinterwordstretchfactor}{4}
\providecommand{\BIBentryALTinterwordspacing}{\spaceskip=\fontdimen2\font plus
\BIBentryALTinterwordstretchfactor\fontdimen3\font minus
  \fontdimen4\font\relax}
\providecommand{\BIBforeignlanguage}[2]{{%
\expandafter\ifx\csname l@#1\endcsname\relax
\typeout{** WARNING: IEEEtran.bst: No hyphenation pattern has been}%
\typeout{** loaded for the language `#1'. Using the pattern for}%
\typeout{** the default language instead.}%
\else
\language=\csname l@#1\endcsname
\fi
#2}}
\providecommand{\BIBdecl}{\relax}
\BIBdecl

\bibitem{van2001refactoring}
A.~van Deursen, L.~Moonen, A.~van Den~Bergh, and G.~Kok, ``Refactoring test
  code,'' in \emph{2nd International Conference on {eXtreme} {Programming} and
  Flexible Processes in Software Engineering}, ser. XP.\hskip 1em plus 0.5em
  minus 0.4em\relax {USA}: {CiteSeer}, 2001, pp. 92--95.

\bibitem{Oliveira2022Lint}
N.~Oliveira, M.~Ribeiro, R.~Bonifácio, R.~Gheyi, I.~Wiese, and B.~Fonseca,
  ``Lint-based warnings in python code: Frequency, awareness and refactoring,''
  in \emph{2022 IEEE 22nd International Working Conference on Source Code
  Analysis and Manipulation (SCAM)}, ser. SCAM 2022, 2022, pp. 208--218.

\bibitem{medeiros2019investigation}
F.~Medeiros, G.~Lima, G.~Amaral, S.~Apel, C.~K{\"a}stner, M.~Ribeiro, and
  R.~Gheyi, ``An investigation of misunderstanding code patterns in c
  open-source software projects,'' \emph{Empirical Software Engineering},
  vol.~24, pp. 1693--1726, 2019.

\bibitem{tahir2016empiricalCat}
A.~Tahir, S.~Counsell, and S.~G. MacDonell, ``An empirical study into the
  relationship between class features and test smells,'' in \emph{APSEC}.\hskip
  1em plus 0.5em minus 0.4em\relax New York, NY, USA: {IEEE}, 2016, pp.
  137--144.

\bibitem{palomba2019smell}
F.~Palomba and A.~Zaidman, ``The smell of fear: On the relation between test
  smells and flaky tests,'' \emph{Empirical Software Engineering}, vol.~24,
  no.~5, pp. 2907--2946, 2019.

\bibitem{hauptmann2013hunting}
B.~Hauptmann, M.~Junker, S.~Eder, L.~Heinemann, R.~Vaas, and P.~Braun,
  ``Hunting for smells in natural language tests,'' in \emph{35th International
  Conference on Software Engineering}, ser. ICSE.\hskip 1em plus 0.5em minus
  0.4em\relax New York, NY, USA: {IEEE}, 2013, pp. 1217--1220.

\bibitem{Fernandes2023Put}
L.~Fernandes, M.~Ribeiro, R.~Gheyi, M.~Delamaro, M.~Guimar\~{a}es, and
  A.~Santos, ``Put your hands in the air! reducing manual effort in mutation
  testing,'' in \emph{Proceedings of the XXXVI Brazilian Symposium on Software
  Engineering}, ser. SBES '22.\hskip 1em plus 0.5em minus 0.4em\relax New York,
  NY, USA: Association for Computing Machinery, 2022, p. 198–207.

\bibitem{hauptmanThesis2016}
B.~Hauptmann, ``Reducing system testing effort by focusing on commonalities in
  test procedures,'' Ph.D. dissertation, Technische Universität München,
  Germany, Jul 2016.

\bibitem{juhnke2021clustering}
K.~Juhnke, A.~Nikic, and M.~Tichy, ``Clustering natural language test case
  instructions as input for deriving automotive testing dsls,'' \emph{Journal
  of Object Technology}, vol.~20, no.~3, pp. 1--14, 2021.

\bibitem{Dalton2020Exception}
F.~Dalton, M.~Ribeiro, G.~Pinto, L.~Fernandes, R.~Gheyi, and B.~Fonseca, ``Is
  exceptional behavior testing an exception? an empirical assessment using java
  automated tests,'' in \emph{Proceedings of the 24th International Conference
  on Evaluation and Assessment in Software Engineering}, ser. EASE '20.\hskip
  1em plus 0.5em minus 0.4em\relax New York, NY, USA: Association for Computing
  Machinery, 2020, p. 170–179.

\bibitem{meszaros2007xunit}
G.~Meszaros, \emph{{xUnit} test patterns: Refactoring test code}.\hskip 1em
  plus 0.5em minus 0.4em\relax Boston, USA: Pearson Education, 2007.

\bibitem{garousi2018smells}
V.~Garousi and B.~K{\"u}{\c{c}}{\"u}k, ``Smells in software test code: A survey
  of knowledge in industry and academia,'' \emph{Journal of Systems and
  Software}, vol. 138, pp. 52--81, 2018.

\bibitem{peruma2019distribution}
A.~Peruma, K.~Almalki, C.~D. Newman, M.~W. Mkaouer, A.~Ouni, and F.~Palomba,
  ``On the distribution of test smells in open source android applications: An
  exploratory study,'' in \emph{29th Annual International Conference on
  Computer Science and Software Engineering}, ser. CASCON.\hskip 1em plus 0.5em
  minus 0.4em\relax USA: {IBM} Corp, 2019, pp. 193--202.

\bibitem{panichella2022test}
A.~Panichella, S.~Panichella, G.~Fraser, A.~A. Sawant, and V.~J. Hellendoorn,
  ``Test smells 20 years later: detectability, validity, and reliability,''
  \emph{Empirical Software Engineering}, vol.~27, no.~7, p. 170, 2022.

\bibitem{Soares2023Replication}
\BIBentryALTinterwordspacing
E.~Soares, M.~Terceiro, N.~Oliveira, M.~Ribeiro, R.~Gheyi, E.~Souza,
  I.~Machado, A.~Santos, B.~Fonseca, and R.~Bonifácio, ``{Manual Tests Do
  Smell! Cataloging and Identifying Natural Language Test Smell - Replication
  Package},'' 7 2023. [Online]. Available:
  \url{http://doi.org/10.6084/m9.figshare.22652620.v2}
\BIBentrySTDinterwordspacing

\bibitem{ubuntuDownload}
C.~Ltd., ``Ubuntu operational system,'' \url{https://ubuntu.com/download},
  [Accessed 02-May-2023].

\bibitem{sparck1972statistical}
K.~Sparck~Jones, ``A statistical interpretation of term specificity and its
  application in retrieval,'' \emph{Journal of documentation}, vol.~28, no.~1,
  pp. 11--21, 1972.

\bibitem{kotrlik2001organizational}
J.~E. Bartlett~II, J.~W. Kotrlik, and C.~C. Higgins, ``Organizational research:
  Determining appropriate sample size in survey research,'' \emph{Information
  technology, learning, and performance journal}, vol.~19, no.~1, pp. 43--50,
  2001.

\bibitem{spacy2023}
\BIBentryALTinterwordspacing
M.~Honnibal and I.~Montani. spacy -- industrial-strength natural language
  processing in python. [Online]. Available: \url{https://spacy.io/}
\BIBentrySTDinterwordspacing

\bibitem{saltzer2020origin}
J.~H. Saltzer, ``The origin of the “mit license”,'' \emph{IEEE Annals of
  the History of Computing}, vol.~42, no.~4, pp. 94--98, 2020.

\bibitem{pennTreebank}
\BIBentryALTinterwordspacing
M.~P. Marcus, B.~Santorini, and M.~A. Marcinkiewicz, ``Building a large
  annotated corpus of {E}nglish: The {P}enn {T}reebank,'' \emph{Computational
  Linguistics}, vol.~19, no.~2, pp. 313--330, 1993. [Online]. Available:
  \url{https://aclanthology.org/J93-2004}
\BIBentrySTDinterwordspacing

\bibitem{universalDependencies}
J.~Nivre, M.-C. de~Marneffe, F.~Ginter, Y.~Goldberg, J.~Haji{\v{c}}, C.~D.
  Manning, R.~McDonald, S.~Petrov, S.~Pyysalo, N.~Silveira, R.~Tsarfaty, and
  D.~Zeman, ``{U}niversal {D}ependencies v1: A multilingual treebank
  collection,'' in \emph{Proceedings of the Tenth International Conference on
  Language Resources and Evaluation ({LREC}'16)}, May 2016, pp. 1659--1666.

\bibitem{honnibal2017spacy}
M.~Honnibal and I.~Montani, ``spacy 2: Natural language understanding with
  bloom embeddings, convolutional neural networks and incremental parsing,''
  \emph{To appear}, vol.~7, no.~1, pp. 411--420, 2017.

\bibitem{van1974foundation}
C.~J. Van~Rijsbergen, ``Foundation of evaluation,'' \emph{Journal of
  documentation}, vol.~30, no.~4, pp. 365--373, 1974.

\bibitem{powers2020evaluation}
D.~M. Powers, ``Evaluation: from precision, recall and f-measure to roc,
  informedness, markedness and correlation,'' \emph{arXiv preprint
  arXiv:2010.16061}, 2020.

\bibitem{soares2020refactoring}
E.~Soares, M.~Ribeiro, G.~Amaral, R.~Gheyi, L.~Fernandes, A.~Garcia,
  B.~Fonseca, and A.~Santos, ``Refactoring test smells: A perspective from
  open-source developers,'' in \emph{Proceedings of the 5th Brazilian Symposium
  on Systematic and Automated Software Testing}, ser. SAST 20.\hskip 1em plus
  0.5em minus 0.4em\relax New York, NY, USA: Association for Computing
  Machinery, 2020, p. 50–59.

\bibitem{soares2022refactoring}
E.~Soares, M.~Ribeiro, R.~Gheyi, G.~Amaral, and A.~Santos, ``Refactoring test
  smells with {JUnit} 5: Why should developers keep up-to-date?'' \emph{IEEE
  Transactions on Software Engineering}, vol.~49, no.~3, pp. 1152--1170, 2023.

\bibitem{NALABS}
\BIBentryALTinterwordspacing
K.~Rajkovic and E.~P. Enoiu, ``Nalabs: Detecting bad smells in natural language
  requirements and test specifications,'' M{\"a}lardalen Real-Time Research
  Centre, M{\"a}lardalen University, Tech. Rep., February 2022. [Online].
  Available: \url{http://www.es.mdu.se/publications/6382-}
\BIBentrySTDinterwordspacing

\bibitem{FEMMER2017190}
H.~Femmer, D.~{Méndez Fernández}, S.~Wagner, and S.~Eder, ``Rapid quality
  assurance with requirements smells,'' \emph{Journal of Systems and Software},
  vol. 123, pp. 190--213, 2017.

\bibitem{Femmer2014}
H.~Femmer, D.~M. Fern\'{a}ndez, E.~Juergens, M.~Klose, I.~Zimmer, and
  J.~Zimmer, ``Rapid requirements checks with requirements smells: Two case
  studies,'' in \emph{Proceedings of the 1st International Workshop on Rapid
  Continuous Software Engineering}, ser. RCoSE 2014.\hskip 1em plus 0.5em minus
  0.4em\relax New York, NY, USA: Association for Computing Machinery, 2014, pp.
  10--19.

\bibitem{aljedaaniAssertionless}
W.~Aljedaani, A.~Peruma, A.~Aljohani, M.~Alotaibi, M.~W. Mkaouer, A.~Ouni,
  C.~D. Newman, A.~Ghallab, and S.~Ludi, ``Test smell detection tools: A
  systematic mapping study,'' in \emph{Evaluation and Assessment in Software
  Engineering}, ser. EASE 2021.\hskip 1em plus 0.5em minus 0.4em\relax
  Association for Computing Machinery, 2021, pp. 170--180.

\end{thebibliography}

\end{document}